\documentclass[pdflatex,sn-mathphys-num]{sn-jnl}

\usepackage{graphicx}%
\usepackage{multirow}%
\usepackage{amsmath,amssymb,amsfonts}%
\usepackage{amsthm}%
\usepackage{mathrsfs}%
\usepackage[title]{appendix}%
\usepackage{xcolor}%
\usepackage{textcomp}%
\usepackage{manyfoot}%
\usepackage{booktabs}%
\usepackage{algorithm}%
\usepackage{algorithmicx}%
\usepackage{algpseudocode}%
\usepackage{listings}%

\theoremstyle{thmstyleone}%
\theoremstyle{thmstyletwo}%

\theoremstyle{thmstylethree}%

\newcommand{\Msun}{\,M_\odot}
\newcommand{\pc} {\,{\rm pc}}
\newcommand{\kpc}{\,{\rm kpc}}

\newcommand{\cm}{\,{\rm cm}}
\newcommand{\yr}{\,{\rm yr}}
\newcommand{\Myr}{\,{\rm Myr}}

\newcommand{\kms}{\,{\rm km\,s^{-1}}}
\newcommand{\K}{\,{\rm K}}
\newcommand{\erg}{\,{\rm erg}}

\newcommand{\Rmax}{R_{\rm max}}
\newcommand{\Reff}{R_{\rm e}}
\newcommand{\fbound}{f_{\rm bound}}
\newcommand{\Mbound}{M_{\rm bound}}
\newcommand{\Mstar}{M_\star}
\newcommand{\Mgas}{M_{\rm gas}}
\newcommand{\fgas}{f_{\rm gas}}
\newcommand{\Mvir}{M_{\rm vir}}
\newcommand{\Rvir}{R_{\rm vir}}

\begin{document}

\title[GCs in cosmological simulations]{Resolved internal dynamics of a globular cluster candidate in a cosmological simulation at cosmic dawn}


\author*[1]{\fnm{Raffaele} \sur{Pascale}}\email{raffaele.pascale@inaf.it}

\author[1]{\fnm{Francesco} \sur{Calura}}\email{francesco.calura@inaf.it}

\author[2]{\fnm{Enrico}\sur{Vesperini}}

\author[2]{\fnm{Alessandro}\sur{della Croce}}

\author[3]{\fnm{Joakim}\sur{Rosdahl}}

\author[4]{\fnm{Alessandro}\sur{Lupi}}


\affil[1]{\orgdiv{Osservatorio di Astrofisica e Scienza dello Spazio di Bologna}, \orgname{INAF}, \orgaddress{\street{Via Gobetti 93/3}, \city{Bologna}, \postcode{40129}, \country{Italy}}}

\affil[2]{\orgdiv{Department of Astronomy, Indiana University}, \orgaddress{\street{Swain West 727 E. 3rd Street}, \city{Bloomington}, \postcode{47405}, \state{IN}, \country{USA}}}

\affil[3]{\orgdiv{Centre de Recherche Astrophysique de Lyon UMR5574}, \orgaddress{\street{Univ Lyon, Univ Lyon1}, \city{Ens de Lyon}, \postcode{F-69230}, \state{Saint-Genis-Laval}, \country{France}}}

\affil[4]{\orgdiv{Como Lake Center for Astrophysics, DiSAT, Università degli Studi dell’Insubria}, \orgaddress{\street{Via Valleggio 11}, \city{Como}, \postcode{IT-22100}, \country{Italy}}}


\abstract{Globular clusters (GCs) are among the oldest and densest stellar systems in the Universe. Their compact sizes and old ages imply they must have assembled rapidly in the early Universe, yet the detailed physical mechanisms responsible for their formation remain uncertain. Although several scenarios have been proposed, cosmological simulations have so far been unable to resolve their detailed formation and internal dynamical structure at birth.  Using a cosmological simulation that uniquely combines sub-parsec resolution with the formation of individual star particles, we follow the gravitational collapse and assembly of a compact stellar system with GC-progenitor mass and size, while directly resolving its internal dynamics and tracing the build-up of its angular momentum. The system assembles most of its stellar mass within $3\Myr$, at $z\simeq10.5$, it rapidly becomes gas-poor, star-dominated, and develops ordered internal rotation acquired largely through the hierarchical assembly of distinct stellar subsystems.  It forms as a satellite of a larger stellar complex within the virial radius of the $\Mvir\simeq1.73\times10^8\Msun$ dark-matter halo hosting it. Its density, size and mass overlap with those of compact stellar systems recently observed by JWST at high redshift. These results indicate that GC progenitors can have well-defined internal kinematic structures already after their formation.} 




\maketitle

\section{Main}
\label{sec:main}

The formation of globular clusters (GCs) remains one of the major unresolved problems in astrophysics. These systems are among the densest stellar structures in the Universe, with typical masses of $10^5$–$10^6\Msun$, half-light radii of a few to tens of parsecs, and are characterized by old stellar populations and little or no dark matter \citep{Baumgardt2020,Baumgardt2021,Gratton2004,Carretta2009,Bastian2018,Milone2022,Moore1996,Conroy2011}. Explaining how such compact, metal-poor and dynamically evolved systems emerge in the early Universe within a self-consistent framework remains a major challenge.


A wide range of formation scenarios has been proposed, including formation within low-mass dark-matter haloes at high redshift \citep{Peebles1984,Boley2009,Trenti2015,Kimm2016,Gutcke2024}, rapid collapse in gas-rich environments under extreme pressures \citep{Elmegreen1997}, fragmentation of dense circumgalactic filaments \citep{vanDonkelaar2026a} or galactic discs \citep{Nipoti2023,Nipoti2024,Mayer2025}, or triggered by rapid halo accretion or galaxy mergers \citep{Renaud2015,Kim2018,Lahen2021,Taylor2025}, and the tidal stripping of larger stellar systems \citep{Bekki2003,Drinkwater2003,Pfeffer2014}. Although these models can reproduce specific properties of GCs, no single mechanism has yet provided a comprehensive explanation for their structural and dynamical characteristics and for the diversity of the GC population \citep{Gratton2012,Bastian2018,Renzini2023}, which may instead result from a combination of these different formation pathways. 

A fundamental limitation of current cosmological simulations is their inability to resolve the internal structure of these systems. In most cases, spatial resolution is insufficient, stars are treated as unresolved populations, and compact stellar systems are sampled with only a limited number of particles. This has so far prevented previous studies from following how such objects form and assemble in a cosmological environment, and how their internal dynamics emerge during this process \citep{Taylor2025}. 

\begin{figure}[h!]
\centering
\includegraphics[width=1\hsize]{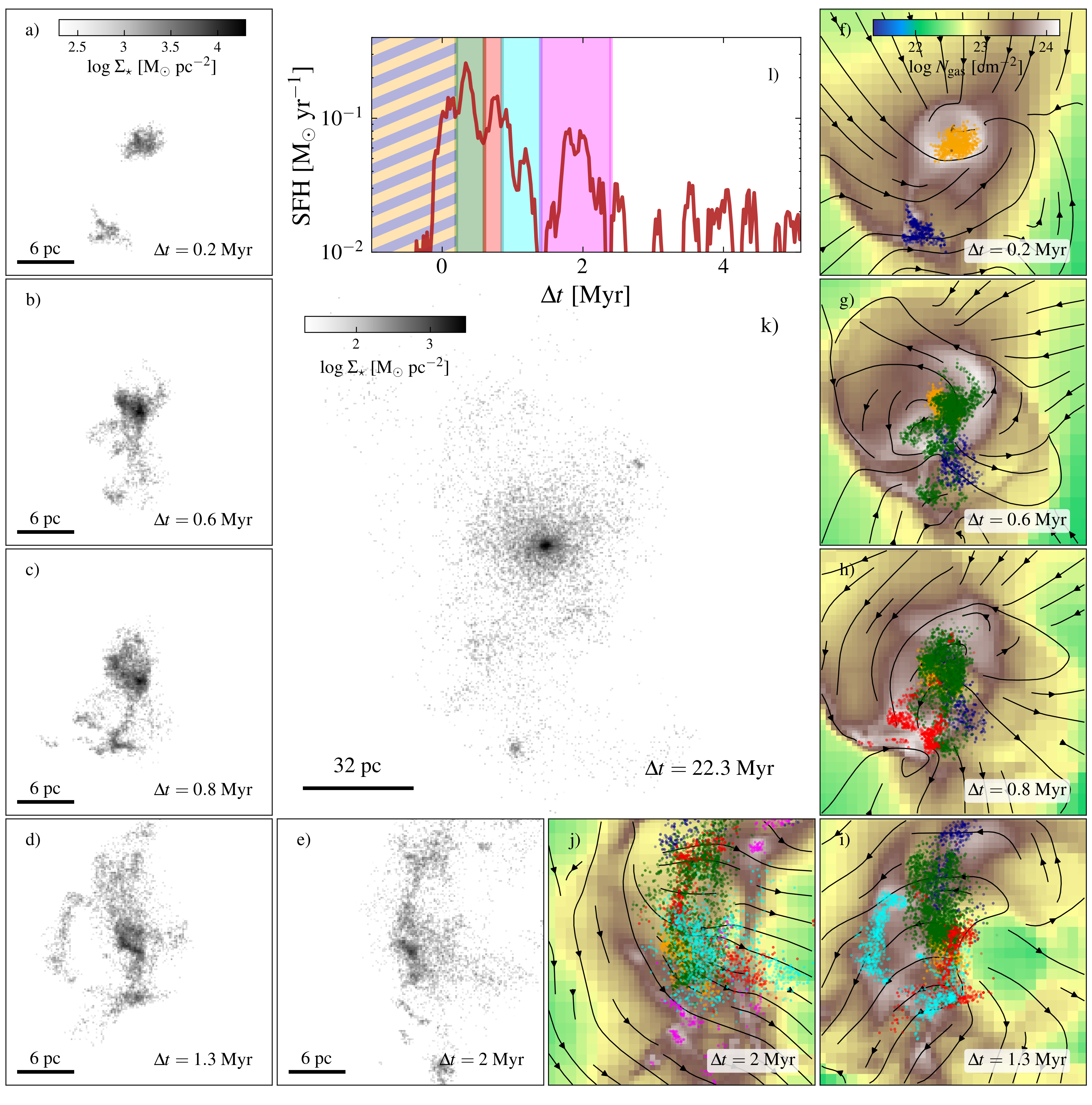}
    \caption{Formation of the GC candidate. Panels (a–e) show the time evolution of the projected stellar distribution, while panels (f–j) show the gas column density in the corresponding regions. Each panel is labelled by the elapsed time since the onset of star formation, $\Delta t$. Panel (l) shows the star formation history (red curve). For the maximum gas densities reached in this regions ($n\simeq10^5$ cm$^{-3}$), the free-fall time can be as small as $t_{\rm ff}\simeq0.15\Myr$, so star formation spans approximately 20 free-fall times. Coloured bands in Panel (l) mark specific time intervals; in panels (f-j), stars formed within each interval are shown using the same colour coding, highlighting how the GC candidate assembles through the collapse of dense gas and the merging of nearby stellar clumps formed at different times. The first coloured bin in the star formation history alternates between orange and blue to distinguish two coeval but spatially distinct stellar components, visible in panel (f). The black arrows show the gas velocity field. Panel (k) shows the stellar distribution of the fully assembled GC candidate. }\label{fig:fig1}
\end{figure} 

Here we overcome these limitations by combining a fully cosmological formation environment with sub-parsec spatial resolution and individual-star sampling using the SIEGE (Simulating the Environment where Globular Clusters Emerged) simulations, a suite of cosmological zoom-in simulations designed to follow the formation of compact stellar systems in a fully self-consistent galaxy formation context \citep{Calura2022,Pascale2023,Calura2024,Pascale2025}. These simulations are evolved down to redshift $z=6.14$, where they reach a spatial resolution of $\simeq0.4\pc$, and follow a dwarf galaxy that, at this redshift, has a virial mass of $4\times10^9\Msun$ (see Section~\ref{subsec:simu}). Each stellar particle represents an individual star drawn from the initial mass function, providing a self-consistent sampling of the stellar mass spectrum and capturing the stochastic nature of stellar feedback. Compact stellar systems with masses of $\sim10^5\Msun$ are therefore resolved with tens of thousands of stellar particles, allowing us to follow their assembly in a fully cosmological context while directly resolving the evolution of their internal structure and kinematics, including the build-up of angular momentum. Further details on the simulation setup are provided in the Methods section. 

Here, we focus on a massive stellar system identified in one of the SIEGE simulations  (see Section~\ref{subsec:simu}) and follow its evolution during the phase in which it rapidly assembles the bulk of its stellar mass. Throughout this work, we refer to this system as a \textit{GC candidate}, as we show that its structural properties closely resemble those of star clusters observed at high redshift and that it shares several structural and dynamical characteristics with local GCs (e.g. size, mass, presence of dark-matter, residual gas, formation time, rotation). Nevertheless, we use the term \textit{candidate} because the subsequent evolution of the system to $z=0$, and hence its direct evolutionary connection to present-day GCs, cannot be established with the present simulation. We discuss these limitations in more detail later.

Figure~\ref{fig:fig1} traces the collapse of dense gas, the emergence of the stellar component and the subsequent assembly of the system from the onset of star formation to the completion of the main burst. In the selected region, star formation begins at redshift $z \simeq 10.5$, with a star formation rate that quickly rises to a peak of $\simeq 0.2\Msun\yr^{-1}$ and remains, on average, around $\sim 0.1\Msun\yr^{-1}$ for approximately $3\Myr$. The GC candidate forms through the rapid hierarchical build-up of several gaseous and stellar structures, some of which collapse into stellar clumps that originate in close proximity and subsequently merge into a single compact stellar system. 

\begin{figure}[h]
\centering
\includegraphics[width=1\hsize]{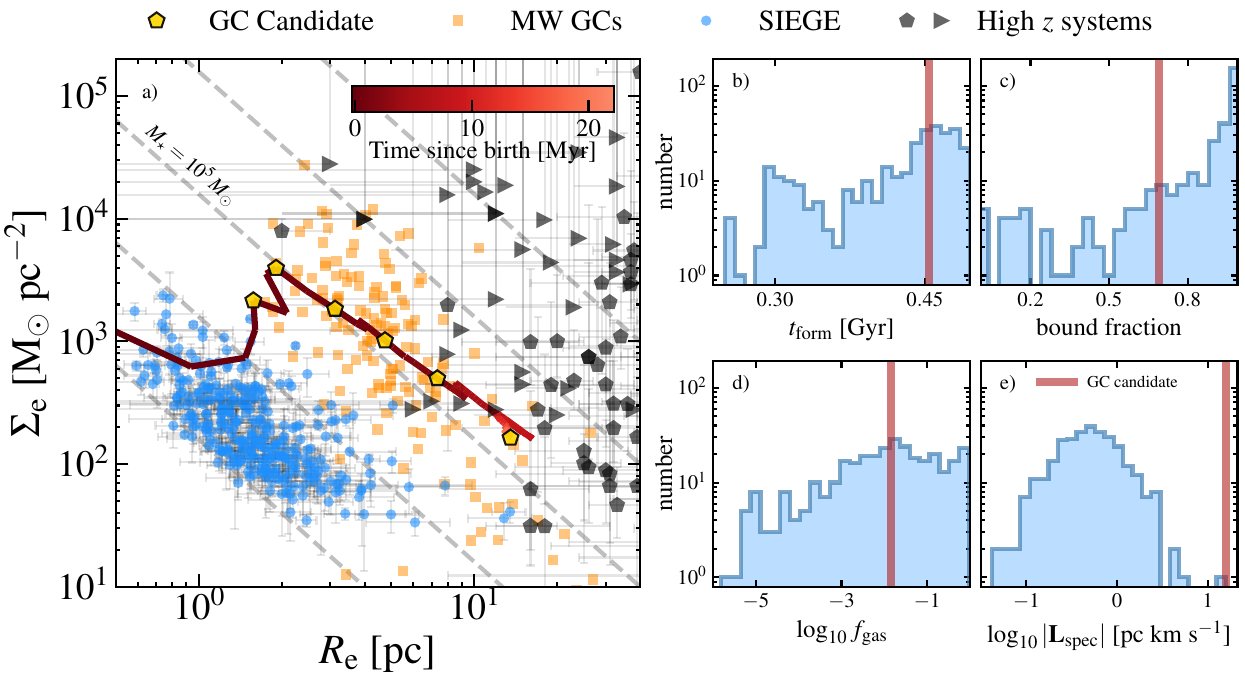}
    \caption{Properties of the GC candidate and of the simulated star-cluster population. Panel a shows the evolution of the GC candidate in the density–size plane. The density is measured through the mean stellar surface density $\Sigma_{\rm e} = M_\star/(2\pi R_{\rm e}^2)$, where $M_\star$ is the total stellar mass, while the size is given by the effective radius $R_{\rm e}$ (see Section~\ref{subsec:strc}). The red curve traces the evolutionary track of the GC candidate, colour-coded by time since the onset of star formation. Yellow markers indicate the specific snapshots shown in Fig.~\ref{fig:fig1}. The panel also includes the other star clusters identified in the SIEGE simulation at $z=10$ in blue, while high-redshift compact stellar systems are shown as black pentagons and triangles. Black triangles represent high-redshift candidates, for which sizes are upper limits and densities are therefore lower limits. Star-cluster measurements from the simulation are averaged over three orthogonal projections, with uncertainties given by the projection-to-projection scatter. Local GCs are shown as orange squares as a reference. Grey dashed lines indicate loci of constant stellar mass. The GC candidate rapidly evolves across this plane during its assembly phase, moving towards the region occupied by the observed high-redshift compact stellar systems. Panels b, c, d, and f show the distributions of formation time, bound fraction, gas mass fraction, and specific angular momentum, respectively, for the simulated star-cluster population (see Section~\ref{subsec:others}). Red vertical lines indicate the position of the GC candidate within each distribution. The high-redshift systems are taken from the catalogue of \cite{Claeyssens2023}.}\label{fig:fig2}
\end{figure}

Prior to the onset of star formation, gas flows coherently into a compact region, building up the natal cloud. Here, turbulent motions remain weak, with typical velocity dispersions of $\simeq4$--$6\kms$, below the sound speed. This natal cloud is associated with a local dark-matter overdensity, which had recently formed through the merger of two low-mass dark-matter structures (see Extended Data Fig.~\ref{fig:extdm}), that may have contributed to the initial accumulation of gas. As discussed later, the natal cloud and its associated dark-matter overdensity are embedded in a more massive dark-matter halo, occupying a region offset from its centre and enclosed within its virial radius. On the scale of the natal cloud the accumulating gas is the dominant component. By the time the first stars form, within a radius of 5 pc (defined as the distance where the gas density has decreased by a factor of 10 from its central value), the gas mass is $M_{\rm gas}\simeq2\times10^5\Msun$ ($\simeq84\%$ of the total mass), compared with a dark-matter mass of $M_{\rm DM}\simeq3.8\times10^4\Msun$. The self-gravity of the accumulated gas therefore drives the subsequent collapse and onset of star formation (see Extended Data Figure~\ref{fig:extfrac}).

The first bursts of star formation occur predominantly within the central, compact region of the natal cloud. Subsequently, further gas accretion and stellar feedback  reshape the natal cloud into a more filamentary and turbulent distribution, within which additional star-forming regions and stellar clumps develop. In all these star-forming regions, the gas reaches densities of $n \simeq10^5\cm^{-3}$, corresponding to free-fall times of $\simeq 0.15\Myr$. On such short dynamical timescales, a large fraction of the available gas is efficiently converted into stars. At the same time, stellar feedback, dominated by stellar winds, efficiently heats the surrounding gas, driving it to high temperatures and reducing its density (see Extended Data Fig.~\ref{fig:phase}). This process rapidly disrupts the dense gas reservoir, leading to a sharp decline in the star formation rate and suppressing any major subsequent episode of star formation after 3 Myr. By the end of this initial star-forming phase, the stellar distribution has already condensed into a compact configuration with a half-mass radius of $\simeq 6\pc$ and has assembled the majority of its stellar mass, reaching $\simeq 1.9 \times 10^5\Msun$, and a central projected stellar surface density of $\sim10^4\Msun\pc^{-2}$. It is resolved with almost $\simeq10^5$ stellar particles, which allows its internal structure and kinematics to be characterised in detail. 

The stellar distribution subsequently expands. This evolution is initially driven by the supervirial state established during star formation and is further enhanced by the rapid removal of residual gas, which weakens the gravitational potential. Most of the increase in size and the associated decrease in surface density occur within the first $\simeq 7-8\Myr$, after which both quantities remain nearly constant. By $\simeq22\Myr$ after the onset of star formation, the GC candidate has effectively completed its assembly phase. The dense gas has been largely exhausted or expelled, leaving behind a compact, gravitationally bound stellar system with a half-mass radius that has increased to $15\pc$, embedded in a locally diffuse environment and no longer experiencing significant gas accretion. 

Recent observations with JWST, aided by strong gravitational lensing, have revealed a population of compact stellar systems at high redshift with extremely high stellar densities and small physical sizes \citep{Claeyssens2023,Messa2024b,Adamo2024,Vanzella2023}. These objects have been proposed as potential progenitors of present-day GCs, although their formation pathways and evolutionary connection to local systems remain uncertain. We follow the early evolution of the simulated GC candidate in the plane defined by stellar density and size, a commonly used diagnostic to characterize these high-$z$ systems. Fig.~\ref{fig:fig2} shows the evolutionary track of the GC candidate over the first $\simeq22\Myr$ after the onset of star formation, and it is compared with both a sample of high-redshift compact stellar systems observed by JWST \citep{Claeyssens2023}, and the population of star clusters identified in the simulation at the same redshift. The GC candidate lies at the massive end of the simulated star-cluster population, which also includes systems with stellar masses of several $10^4\Msun$, and is the largest compact stellar system present in the simulation at this redshift. It rapidly evolves across this parameter space during its early assembly phase, reaching stellar densities and sizes consistent with those inferred for observed high-redshift compact systems. In particular, it matches the better-resolved high-$z$ systems at larger sizes and lower densities, where the comparison is more robust and also less affected by potential biases associated with lensing magnification \citep{Claeyssens2026}. However, one caveat to this agreement is that the observed radii are half-light radii, which at the highest redshifts primarily trace the rest-frame UV emission, whereas the simulations provide stellar half-mass radii. Spatial variations in stellar populations and mass-to-light ratios may therefore introduce additional differences between the two quantities \citep{Claeyssens2023,Ambachew2022}. We note that exceptional systems such as the Cosmic Gems \citep{Adamo2024} clusters at ($z\simeq9.625$), detected thanks to exceptionally high gravitational lensing magnifications and characterized by effective radii down to sub-parsec scales and stellar surface densities exceeding $10^5\Msun\pc^{-2}$, fall outside the plotted range, which focuses on the broader observed population. Fig.~\ref{fig:fig2} also presents present-day GCs as a useful reference for the properties of old stellar systems, although the young age of the simulated star clusters makes high-redshift compact stellar systems the most direct observational comparison; we discuss the possible long-term evolution of the GC candidate later in the manuscript. 

\begin{figure}[h]
\centering
\includegraphics[width=1\hsize]{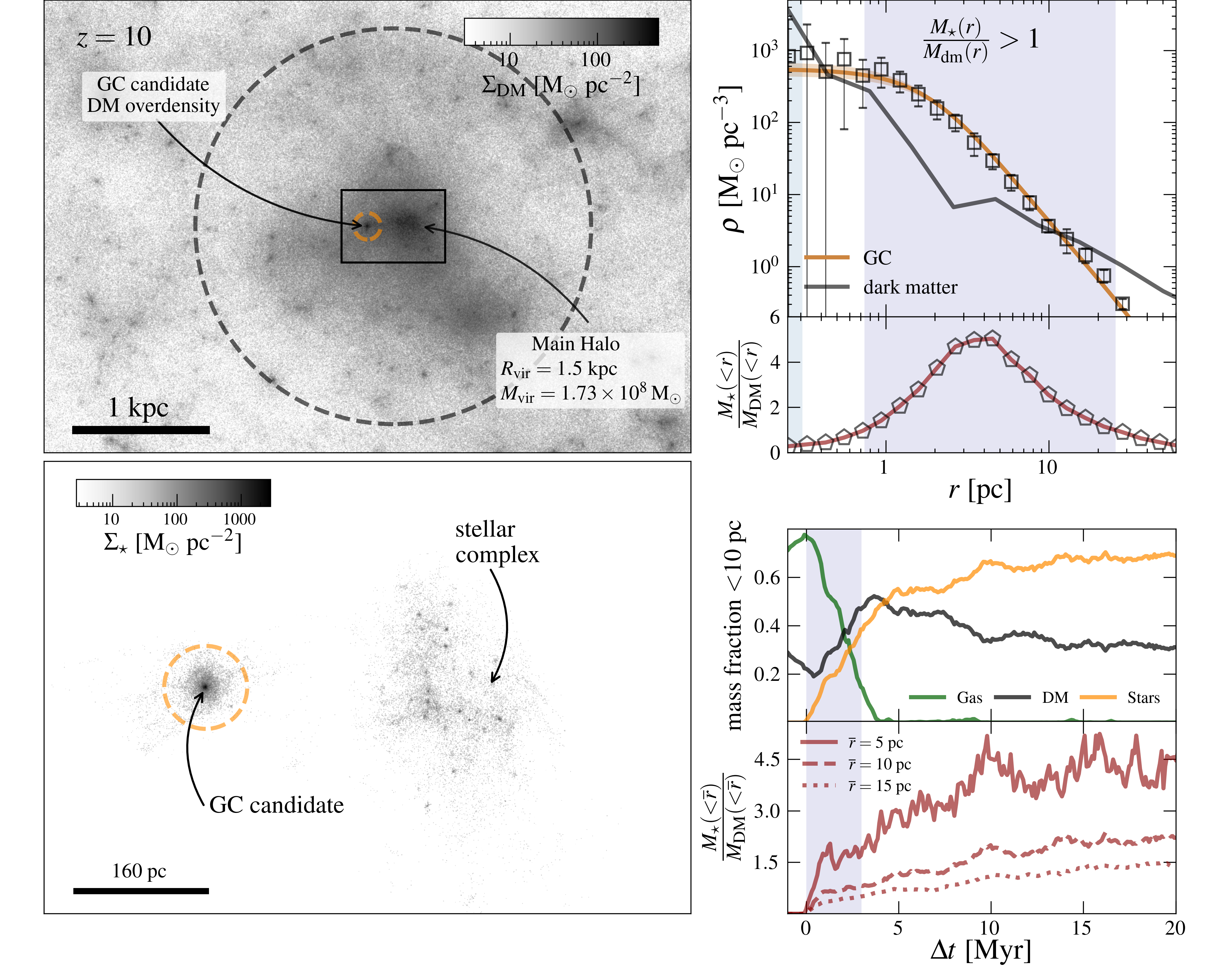}
\vspace{-0.5cm}
\caption{Top left panel: projected dark matter density map showing the filamentary structure hosting the GC candidate. The region marked by a rectangle is shown as a zoom in the bottom-left panel, which displays the corresponding projected stellar density. Here, the GC candidate is visible on the left, while a nearby stellar complex lies to the right. The stellar complex is embedded within a dark matter halo with virial radius of 1.5 kpc, shown in the top panel by a black dashed circle, and virial mass $1.73\times10^8\Msun$. The dashed orange circle in the bottom left panel indicates a radius of $\simeq50\pc$, approximately twice the region within which stars dominate over dark matter. Top right panels: radial density profiles of dark matter (grey line) and stars (points with error bars). The bottom-right panel shows the radial profile of the stellar-to-dark-matter mass ratio (red curve and points). The light blue shaded region to the left marks the spatial resolution limit (cell size), below which gravity is not reliably resolved. The darker blue region highlights where the stellar-to-dark-matter mass ratio exceeds unity. The orange band represents the best-fitting King \cite{King1966} model. Bottom right panels: Mass fractions within 10 pc from the GC candidate centre as a function of time since the onset of star formation for stars (orange), gas (green), and dark matter (black), and stellar-to-dark-matter mass ratio within three characteristic apertures (red). The shaded blue band spans the $3\Myr$ period of star formation.}\label{fig:dm}
\end{figure}

The spatial distribution of stars and dark matter at the end of the main assembly phase is shown in Fig.~\ref{fig:dm}. At this stage, the system is gas-free and no ongoing star formation is present. The GC candidate lies within the virial radius ($\Rvir\simeq1.5\kpc$) of a massive ($\Mvir\simeq1.73\times10^8\Msun$) dark matter halo (highlighted in Fig.~\ref{fig:dm}) at a distance of $\simeq0.2\Rvir$ from its centre. This massive halo is itself embedded within a larger cosmic filament and hosts a complex of lower-mass stellar systems populating its central regions. Although a local dark matter overdensity is still present at the location of the GC candidate, regions extending out to about 1.5$\Reff\simeq25$ pc receive the largest contribution from the stellar component, with the stellar distribution significantly more centrally concentrated, while the surrounding dark matter remains diffuse and extends over much larger spatial scales. Within this radial extent, the stellar-to-dark-matter mass ratio reaches values of $\sim 5$ (top right panels of Fig.~\ref{fig:dm}). Throughout the whole formation and assembly process, dark matter never accounts for the largest fraction of the enclosed mass: gas dominates before and during the main star-formation episode, and is progressively overtaken by stars. The dark matter contribution gradually decreases with time as a consequence of tidal stripping by the host galaxy. This process remains continuous throughout the entire time interval analysed, preferentially removing loosely bound material from the outer regions of the GC candidate, where the cluster transitions into the surrounding ambient dark matter distribution (see Extended Data Figure~\ref{fig:extfrac}). Further analysis of the stripping of loosely bound matter will be presented in a future work. 

\begin{figure}[!t]
    \centering
    \includegraphics[width=1\hsize]{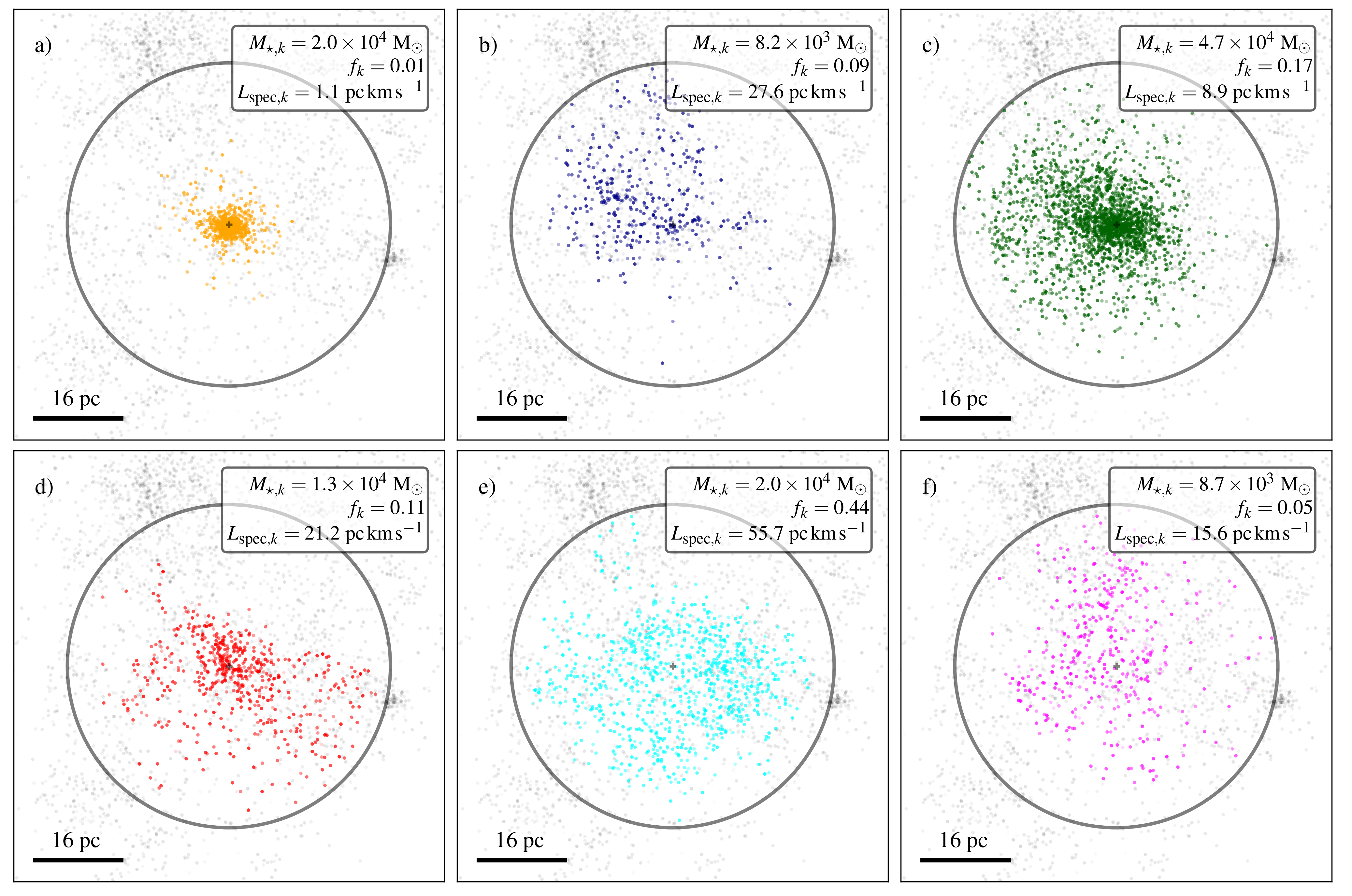}
    \caption{Internal kinematics of the GC candidate. Panels (a-f) show the stellar distributions at $\simeq22$ Myr for stars formed at different times, using the same intervals and colours as in Fig.~\ref{fig:fig1}. For each component $k$, $M_{\star,k}$ is the mass of that component, $f_k$ gives its fractional contribution to the final angular momentum (counter-aligned components can have negative $f_k$), $L_{\rm spec, k}$, the corresponding specific angular momentum. The grey circle marks 30 pc from the centre: only stars lying within this radius at $\simeq22\Myr$ and formed within the first $\simeq2.5\Myr$ are used for this decomposition and in the computation of $f_k$, $M_{\star,k}$ and $J_{z,k}$. In all panels, the line of sight is aligned with the angular-momentum vector.}
    \label{fig:fig4}
\end{figure}

A key advantage of the adopted star-by-star modelling is the ability to resolve the internal structure and kinematics of compact stellar systems without being limited by poor particle sampling. In particular, we can follow the emergence of rotation and the build-up of angular momentum in the GC candidate throughout its assembly, probing the viability of the underlying formation mechanisms down to small mass scales and in a cosmological context. These properties are particularly informative because they can encode the kinematics of the progenitor gas and the dynamical processes involved in cluster assembly, as well as the influence of the surrounding tidal environment \citep[e.g.][]{vanDonkelaar2026b}. Moreover, the eventual rotation established during this phase provides a dynamically important initial condition for the cluster's subsequent evolution \citep{Ernst2007}, even if its imprint is progressively modified over a GC lifetime \citep{Bianchini2026}.

Fig.~\ref{fig:fig4} provides a view of the internal structure and kinematics of the GC candidate after $\simeq22\Myr$ by decomposing it into stellar populations according to their formation time as defined in panel l) of Fig.~\ref{fig:fig1}. The populations associated with approximately the first three bursts (panels a--c) form close in time and remain preferentially concentrated in the centre of the natal cloud; together, they account for $\simeq50\%$ of the stellar mass within 30 pc. The populations formed at later times, associated with the surrounding filamentary environment, are progressively dispersed, with their stars phase-mixing predominantly into the outer regions of the GC candidate. This decomposition reveals large differences in how individual populations contribute to the angular-momentum budget. One of the populations formed at later times alone contributes $\simeq44\%$ of the total angular momentum and has the highest specific angular momentum. This population forms in a surrounding gaseous filament and initially orbits the main forming structure, suggesting that part of its orbital angular momentum is converted into the internal spin of the assembled system. In comparison, one of the populations associated with the first bursts is the largest contributor to the stellar mass but accounts for $\simeq20\%$ of the total angular momentum, and has one of the lowest specific angular momenta. 

\begin{figure}
    \centering
    \includegraphics[width=1\hsize]{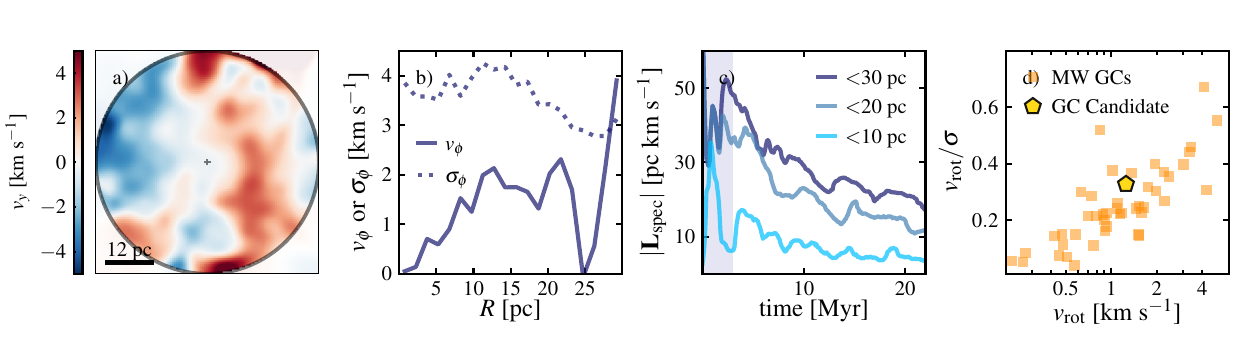}
    \caption{Panel (a) shows the stellar velocity field, colour-coded by the $v_y$ component, to highlight rotation. The spatial scale is the same as in Fig.~\ref{fig:fig4}. Panel (b) shows the radial profiles of mean azimuthal velocity $v_\phi$ and dispersion $\sigma_\phi$ at $\simeq22$ Myr. Panel (c) shows the specific angular momentum of the GC candidate computed using stars within fixed apertures of 10, 20, and 30 pc as a function of time. Panel (d) shows rotation velocity, $v_{\rm rot}$, versus $v_{\rm rot}/\sigma_\phi$ (see Section~\ref{subsec:rot}). MW GCs are shown for comparison, following the procedure of \cite{vanDonkelaar2026b}: $v_{\rm rot}$ values are taken from \cite{Sollima2019}, while $\sigma_\phi$ values from \cite{Vasiliev2021} are interpolated at the half-mass radius. GCs with $R_{\rm GC}>3\kpc$ are excluded.}
    \label{fig:fig5}
\end{figure}

\begin{figure}
    \centering
    \includegraphics[width=1\hsize]{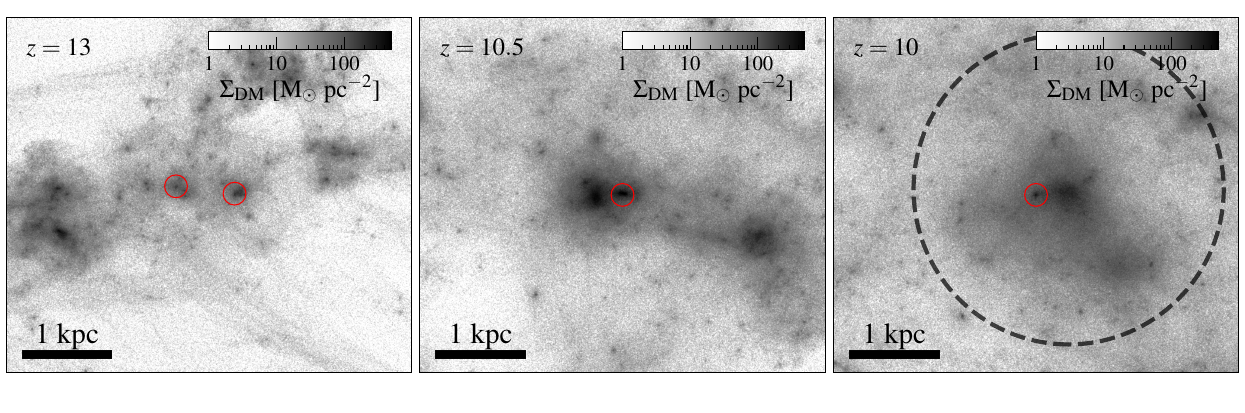}
    \caption{Dark-matter distribution around the GC candidate at three redshifts. The right panel shows the system at the end of the time span considered in the manuscript ($z=10$), the same as in the top left panel of Fig.~\ref{fig:dm}. The extended dark-matter distribution corresponds to the main halo, whose virial radius ($\Rvir=1.5\kpc$) is marked by the grey dashed circle, while the red circle identifies the smaller dark-matter overdensity hosting the GC candidate. The central panel shows the system at $z=10.5$, approximately at the onset of star formation in the GC candidate. The left panel shows, at $z=13$, the two distinct low-mass dark-matter structures that subsequently merge into the single overdensity hosting the GC candidate.}
    \label{fig:extdm}
\end{figure}

\begin{figure}[!t]
    \centering
    \includegraphics[width=1\hsize]{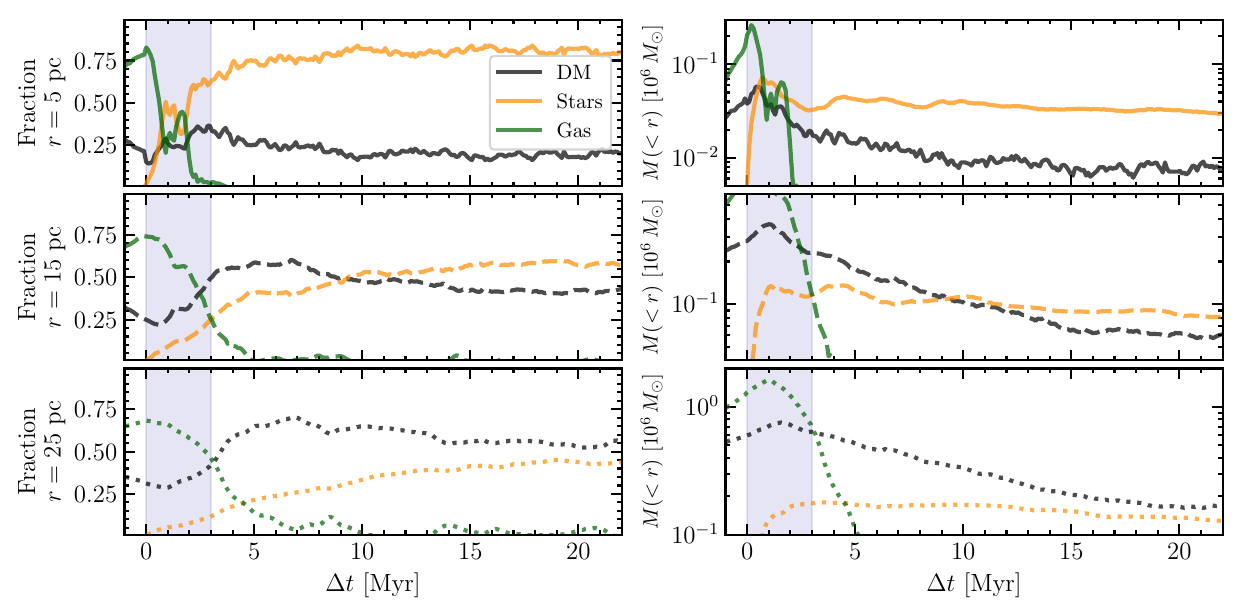}
    \caption{Enclosed gas, stellar, and dark-matter mass fractions within characteristic apertures centred on the natal cloud / GC candidate. Negative times precede the onset of star formation ($\Delta t=0$). The central region is initially gas dominated, while gas is progressively converted into stars during the subsequent star-formation episode. Right panels: same as the left panels, but showing the masses. The blue band marks shows the $3\Myr$ episode of star formation. We note that the enclosed masses include both gravitationally bound and unbound particles; this is especially relevant for dark matter at 15 and 25 pc, where the amount actually bound to the GC candidate is lower than the value shown.}
    \label{fig:extfrac}
\end{figure}

The total specific angular momentum, measured within $10$, $20$, and $30$ pc, increases rapidly during the main assembly phase (see Fig.~\ref{fig:fig5}), reaches a maximum shortly after the system is fully formed, and subsequently declines gradually, indicating a progressive redistribution of angular momentum. We explored the connection between the angular momentum of the natal gas and that established in the first stellar populations, comparing the angular momentum of the pre-collapse gas with that of the stellar component during the early assembly phase. We find that the angle between the two angular-momentum vectors remains $\simeq30^\circ$ during the first $\simeq1\Myr$, while their magnitudes are comparable, suggesting that the first stellar populations partially inherit the angular momentum of the natal gas. This alignment is subsequently lost as the later, more extended stellar populations assemble. Taken together, in a fully cosmological context, these results indicate that the angular momentum of the GC candidate has a composite origin, with different contributions acting at different stages of its formation, and that hierarchical assembly plays a key role in establishing its internal angular momentum even on these very small mass scales.

The radial profiles at $\simeq22\Myr$ reveal clear differential rotation: the inner regions are dominated by velocity dispersion, with $\sigma_\phi$ exceeding the rotation velocity, while at larger radii random motions become increasingly less important. The projected stellar velocity field, viewed along the angular momentum vector, shows a coherent large-scale rotational pattern across the system. Remarkably, already at the end of its assembly phase, the GC candidate lies on the sequence populated by present-day Milky Way GCs in the $v_{\rm rot}/\sigma$--$v_{\rm rot}$ plane \citep{Sollima2019,Vasiliev2021}, showing both a rotation velocity and a level of rotation relative to random stellar motions comparable to those of local GCs. A similar agreement is found for some high-redshift clusters in the cosmological simulations of \cite{vanDonkelaar2026b}. 


The resolved formation of a GC candidate presented here highlights the emergence of the structural and dynamical primordial properties of these systems. Such a direct characterization of a forming GC candidate is possible only thanks to cosmological simulations that simultaneously achieve sub-parsec resolution and model stars individually, allowing the internal dynamical structure of the system to be resolved during its assembly phase. In the simulation, the GC candidate forms through a rapid and localized collapse within a dark-matter overdensity, following the merger of two very lower-mass progenitors. During the collapse and on the scales of the forming GC candidate, the gas dominates the enclosed mass and gravitational acceleration. The system assembles most of its stellar mass within only a few Myr, through multiple closely spaced episodes of star formation. The first stellar populations form from the initial collapsing cloud, while subsequent stellar components also form in nearby filamentary structures generated by feedback-driven gas compression. Star formation is eventually quenched by the rapid consumption of the dense gas reservoir, aided by stellar winds, before the onset of the first supernovae \citep{Calura2022}. This naturally produces a gas-poor, compact and gravitationally bound stellar system within only a few Myr, without requiring a dominant dark-matter component on cluster scales, consistent with the properties inferred for compact JWST sources at high redshift. The formation process is consistent with pathways identified in recent simulations of massive star-cluster formation. \cite{Sameie2023} similarly find rapid cluster formation from dense gas reaching comparable surface densities, albeit at lower spatial resolution, while \cite{Garcia2023} find compact clusters forming from the collapse of self-gravitating clouds, but at substantially lower masses. GC-like systems have also been found to form within low-mass dark-matter haloes \citep[e.g.][]{Kimm2016,Gutcke2024}, but these systems either retain a dominant dark-matter component after formation \citep{Kimm2016} or have substantially lower stellar masses \citep{Gutcke2024}.

The subsequent long-term evolution of the GC candidate remains uncertain. Our simulation does not follow the system to low redshift and therefore cannot establish its fate at the present day. By the final simulation snapshot at $z=6.14$, the cluster has become a satellite of the main galaxy in the simulated region. Its long-term evolution will depend on its orbital history, the evolving tidal field of the growing host galaxy \citep{Renaud2011,LiGnedin2019}, and collisional processes \citep{DellaCroce2026} that are not captured here, which can affect its internal dynamics, for instance by reducing its rotation \citep{Bianchini2026}. Nevertheless, collisional $N$-body calculations including stellar evolution, two-body relaxation and an external tidal field indicate that a cluster of this mass can survive for more than a Hubble time under Milky-Way-like tidal conditions, supporting its identification as a plausible progenitor of a present-day GC \citep{DellaCroce2026}.

Our results demonstrate that the formation of a compact, rotating and dynamically structured GC candidate and its internal dynamics can be captured directly in cosmological simulations when both parsec-scale gas collapse and the individual stellar population are resolved, and that some of the structural and kinematic properties of GCs may already be established at birth, such as the presence of ordered rotation shortly after formation. Their internal kinematics may therefore preserve a fossil record of the physical conditions in their birth environment, although the extent to which this information survives depends on the subsequent collisional evolution. Although our conclusions are based on a single object, extending this approach to larger volumes and more massive environments will open the way to population-level studies of GC formation and internal dynamics.

\section{Methods}
\label{sec:methods}

\subsection{Simulation set-up}
\label{subsec:simu}

The simulation analysed in this work is part of the SIEGE (Simulating the Environment where GCs Emerged) project, a suite of cosmological zoom-in hydrodynamical simulations designed to follow the formation of compact stellar systems within a fully self-consistent galaxy formation framework. The project comprises four simulations that differ in the implementation of specific aspects of the star formation and/or stellar winds model, as the adoption of a top heavy IMF, or the way feedback from massive stars is implemented. The simulations and their physical modelling are described in detail in \cite{Calura2022,Calura2024}, while their application to the formation of compact stellar systems has been explored in \cite{Pascale2023,Pascale2025}. 

The simulation is performed with the adaptive mesh refinement code \textsc{RAMSES}~\cite{teyssier2002} in a cosmological box of side $5\,h^{-1}$ Mpc (comoving), with a zoom-in region centred on a dwarf-galaxy halo. A flat $\Lambda$CDM cosmology is adopted, with $\Omega_m = 0.276$ and $H_0 = 70.3\kms$ Mpc$^{-1}$~\cite{omori2019}. Dark matter is resolved with particles of mass $165\Msun$, and the simulation reaches a spatial resolution of $\simeq 0.3\,h^{-1}$ pc at $z = 6.14$. The target halo has a virial mass of a few $10^9\Msun$ at $z = 6.14$, corresponding to the progenitor of a present-day dwarf galaxy.

Gas cooling includes hydrogen, helium and metal-line processes under the assumption of equilibrium thermochemistry~\cite{few2014}, together with a redshift-dependent ultraviolet background~\cite{haardt1996}. Star formation occurs in cold ($T < 2 \times 10^4\K$), dense gas at the highest refinement level, following a Schmidt-Kennicut like relation of the form
\begin{equation}\label{for:sfe}
    \dot{\rho}_\star = \epsilon_{\rm SFE}\frac{\rho_{\rm gas}}{t_{\rm ff}},
\end{equation}
where $t_{\rm ff}$ is the local free-fall time and $\epsilon_{\rm SFE}$ is the star formation efficiency per free-fall time. In this work, we adopt a high-efficiency model with $\epsilon_{\rm SFE}=1$. This choice is motivated by numerical and observational studies showing increasingly efficient star formation in the dense environments associated with massive cluster formation, with efficiencies reaching high values and, in the densest simulated systems, approaching unity \citep{Grudic2019,Polak2024,Turner2015,RicoVillas2020} (see also discussion in Section~\ref{subsec:gp}).

Star formation occurs above a redshift-dependent density threshold, $\rho_{\rm thr}=m_\star/[0.9(\Delta x)^3]$, where $m_\star=32\,M_\odot$ is a reference base stellar particle mass and $\Delta x$ is the physical size of a maximally refined cell \cite{Yaghoobi2022,Calura2022}. At $z\simeq10$--$10.5$, this corresponds to $\simeq2\times10^4$ cm$^{-3}$. The high threshold is enabled by the sub-pc resolution of SIEGE, which allows gas collapse to be followed to high densities before star formation occurs \cite{Hopkins2013,Kimm2016}.
A key feature of the simulation is that stars are formed individually: each stellar particle represents a single star drawn from a Kroupa initial mass function in the mass range $0.1$--$100\Msun$, following the method described in Sormani et al.~\cite{sormani2017}. This enables a self-consistent sampling of the stellar mass spectrum and captures the intrinsically stochastic nature of stellar feedback, particularly in low-mass systems where individual massive stars and supernovae can dominate the energy budget. This approach is essential at the achieved resolution, where individual gas cells often contain only enough mass to form a few stars, making discrete sampling of the stellar population necessary to accurately represent both star formation and feedback.

Stellar feedback is implemented at the level of individual stars. Stars in the mass range $8$--$40\Msun$ contribute to feedback through both stellar winds and Type II supernova explosions, while stars in the mass range $1.5$--$8\Msun$ only through stellar winds. During the pre-supernova phase, stars inject mass and energy continuously into the surrounding medium through stellar winds, while at the end of their lifetime they explode as supernovae, releasing $10^{51}\erg$ of thermal energy. This discrete and time-dependent treatment of feedback allows us to capture the stochastic impact of individual massive stars on the surrounding gas, which is particularly important during the early phases of cluster formation.

The adopted numerical framework has been extensively tested in previous works of the SIEGE project, which explored the impact of star formation efficiency and stellar feedback modelling on the formation of dense stellar systems and high-redshift star-forming clumps. In particular, the present $\epsilon_{\rm SFE}=1$ realization produces the densest and most compact stellar systems, whereas lowering the efficiency to $\epsilon_{\rm SFE}=0.1$ results in substantially more extended and less dense systems \cite{Calura2024}. The adopted pre-SN feedback also affects the gas morphology: the continuous injection of mass and energy through stellar winds over the lifetime of massive stars preserves a more structured and filamentary cold-gas distribution than a different implementation explored in the project, in which the pre-SN contribution is injected impulsively at stellar birth. Adopting a top-heavy IMF also results in substantially more diffuse stellar systems. None of the other SIEGE realizations forms systems with properties comparable to the most compact population found in the $\epsilon_{\rm SFE}=1$ simulation, motivating our focus on this realization.

As mentioned, the simulation produces a population of compact stellar systems with masses $10^3$--$10^5\Msun$, sizes of $1$--$3\pc$, and surface densities up to $\sim10^4\Msun\pc^{-2}$. In this work, we focus on the most massive system, which is resolved with tens of thousands of stellar particles, enabling a detailed characterization of its internal structure and kinematics at formation.

The evolution of the GC candidate is followed through a sequence of closely spaced simulation snapshots during its main assembly phase. Starting from redshift $z=10.5$, snapshots are separated by time intervals of order $\sim 0.11$ Myr, for a total of $22.3\Myr$, allowing the collapse of the gas, the build-up of the stellar component and the emergence of the internal structure and dynamic to be resolved on timescales comparable to the local dynamical time of the system.

\subsection{Characterization of the GC candidate}
\label{subsec:gcprop}

In this section, we describe the methods used to derive some of the structural and kinematic properties of the GC candidate presented throughout the manuscript. These quantities are used in several of the figures and analyses discussed in the main text.

\subsubsection{Centre determination and projected half-mass radius}
\label{subsec:strc}

Here we briefly describe the numerical methods adopted to compute the half-mass radii and the dynamical properties of the GC candidate. These quantities are evaluated independently at each simulation snapshot, after recentering the system. The centre of the GC candidate is computed using the shrinking-sphere method introduced by \cite{Power2003}. Starting from an initial spherical region enclosing the stellar distribution, the centre of mass is iteratively recomputed while progressively shrinking the radius of the sphere. The procedure is halted when the innermost sphere contains 100 particles, whose centre of mass is assumed to define the final GC candidate centre. This approach is robust against asymmetries and contamination from loosely bound or external particles. After recentering, the half-mass radius is defined as the projected distance from the GC candidate centre enclosing half of the stellar mass contained within a maximum projected radius of 50 pc. We show this and derived quantities as measured along a single projection, but we quantified their dependence on projection by repeating the calculation along three orthogonal projections throughout the evolution of the GC candidate. For instance, the tracks in the density-ize plane (Fig.~\ref{fig:fig2}) are very similar: deviations in half-mass radius peak at approximately 20\% during the initial formation phase, subsequently remain below 10\%, and have time-averaged absolute deviations from the mean of 4.8-7.8\%, depending on the projection. Variations in stellar surface density are larger, primarily because of its quadratic dependence on the projected half-mass radius, while the enclosed stellar mass is identical across projections. Overall, projection-induced variations are therefore small and do not affect the inferred structural evolution of the GC candidate. The choice of $R_{\rm max}=50$ pc is motivated by the need to minimise contamination. A posteriori, this value corresponds to approximately three times the typical half-mass radius measured for the GC candidate, thus ensuring that the bulk of its stellar component is included in the calculation.

\subsubsection{Angular momentum and rotational quantities}
\label{subsec:rot}

To compute the angular momentum shown in Fig.s~\ref{fig:fig4} and~\ref{fig:fig5}, we proceed as follows. For every snapshot, we consider stars enclosed within three characteristic radii, corresponding to $10$ pc, $20$ pc, and $30$ pc. The velocities of the selected particles are referred to the cluster systemic velocity, computed as the average velocity of the stellar sample. The total angular momentum vector is then calculated as

\begin{equation}
\mathbf{L}_{\rm tot} = \sum_k m_k (\mathbf{r}_k \times \mathbf{v}_k),
\end{equation}
while the specific angular momentum is
\begin{equation}
\mathbf{L}_{\rm spec} = \frac{\mathbf{L}_{\rm tot}}{M_{\rm tot}},
\end{equation}
where $m_k$, $\mathbf{r}_k$, and $\mathbf{v}_k$ are the mass, position vector, and velocity vector of the $k$-th stellar particle relative to the cluster centre and systemic velocity, while $M_{\rm tot}$ is the total stellar mass within the aperture. The direction of $\mathbf{L}_{\rm tot}$ defines the orientation of the GC candidate rotation axis, whereas its magnitude provides a measure of the overall rotational strength of the system. 

After defining the rotation axis
\begin{equation}
\hat{\mathbf{l}} = \frac{\mathbf{L}_{\rm tot}}{|\mathbf{L}_{\rm tot}|},
\end{equation}
the rotational velocity of each particle around this axis is computed as
\begin{equation}
v_{\phi,k} = \frac{\mathbf{L}_k \cdot \hat{\mathbf{l}}}{m_k R_k},
\end{equation}
where $\mathbf{L}_k$ is the angular momentum of the $k-$th particle and $R_k$ is its cylindrical distance from the rotation axis,
\begin{equation}
R_k = \left\lVert \mathbf{r}_k - (\mathbf{r}_k \cdot \hat{\mathbf{l}})\hat{\mathbf{l}} \right\rVert.
\end{equation}

In Fig.~5, $v_{\rm rot}$ is computed as the mass-weighted mean azimuthal velocity, $v_\phi$, measured with respect to the final angular-momentum direction, while $\sigma_\phi$ is the corresponding mass-weighted azimuthal velocity dispersion. Both quantities are measured using all stars within 30 pc of the GC candidate centre, corresponding to approximately $2R_{\rm e}$ at 22 Myr. For comparison, Milky Way GC maximum rotation velocities are taken from \cite{Sollima2019}, with velocity dispersions from \cite{Vasiliev2021} evaluated at the projected half-light radius; clusters associated with the population population of the galactic Bulge ($R_{\rm GC}<3$ kpc) are excluded.

\subsection{Characterization of the SIEGE star cluster population}
\label{subsec:others}

Here, we describe the methods used to characterize the entire SIEGE star cluster population shown in Fig.~\ref{fig:fig2}. We consider the same star clusters identified in \cite{Pascale2025} at $z=10$, corresponding to the end of the $\simeq22$ Myr interval over which we follow the evolution of the GC candidate. \cite{Pascale2025} provided a first characterization of these systems, including their formation times, stellar masses, projected half-mass radii $\Reff$, and Sérsic indices $n$, with the latter three quantities derived by fitting their projected stellar density distributions with Sérsic profiles along three independent projections. We now extend this characterization and describe how the bound stellar mass fraction, gas fraction, and specific angular momentum are derived for the present analysis.

\subsubsection{Bound stellar mass fraction}
\label{subsec:fbound}

To determine the bound stellar mass fraction, we adopt the mean $R_{\rm e}$ and $n$ across the three projections and define $\Rmax$ as the radius enclosing 99\% of the stellar mass of the resulting Sérsic profile. We then use $\Rmax$ as a spherical aperture to select the stellar particles associated with each star cluster.

We identify the self-bound stellar component through an iterative energy-based procedure. The systemic velocity is first computed as the mass-weighted mean velocity of the stars within $\Reff$ and is kept fixed throughout the calculation. For each star within $\Rmax$, we compute the specific kinetic energy in this reference frame and the gravitational potential generated by the other stellar particles within the same aperture. The potential is evaluated using a Plummer-softened potential with $\epsilon\simeq0.31$ pc, corresponding to the finest physical resolution of the simulation at this epoch. Stars with positive total specific energy are removed, the stellar potential is recomputed using only the remaining particles, and the procedure is iterated until the set of bound particles no longer changes. The bound stellar mass fraction is then defined as $\fbound\equiv \Mbound(<\Rmax)/\Mstar(<\Rmax)$, where the denominator is the total stellar particle mass included in the initial $\Rmax$ selection. We normalize $\fbound$ to the stellar particle mass within $\Rmax$, rather than to the stellar mass inferred from the Sérsic fit, so that both numerator and denominator are defined from the same initial particle selection. The particle- and Sérsic-based mass estimates are nevertheless consistent within the adopted apertures, with differences that in most cases remain within the projection-to-projection scatter among the Sérsic fits.

The gravitational contributions of gas and dark matter are neglected in this calculation, as both provide only a minor contribution to the gravitational potential on the scales of the clusters. In particular, we exclude dark matter because our aim is to determine whether these systems are self-bound by their stellar gravitational potential alone. We tested the dependence of the inferred bound stellar mass on the adopted aperture by repeating the full unbinding calculation within $2\Rmax$. The resulting bound stellar masses are well converged, with differences of order a few per cent, indicating that $\Rmax$ is sufficiently large to capture the self-bound stellar component.

\subsubsection{Gas mass fraction}
\label{subsec:gas}

We measure the gas mass within a sphere of radius $\Rmax$ centred on each cluster, and define the gas mass fraction as
\begin{equation}
    \fgas(<\Rmax) = \frac{\Mgas(<\Rmax)}{\Mgas(<\Rmax)+\Mstar(<\Rmax)}.
\end{equation}
For the very small number of systems with particularly extended S\'ersic profiles, for which $\Rmax>10\Reff$, we limit the aperture used for the $\Rmax$ gas measurement to $3\Reff$ to reduce contamination from gas associated with the surrounding environment.

\subsubsection{Specific angular momentum}
\label{subpop}

We compute the total stellar angular momentum as described in Section~\ref{subsec:rot}, with the difference that here the sum is restricted to the stellar particles identified as bound by the unbinding procedure described above. The specific angular momentum is then obtained by dividing the total angular momentum by the corresponding bound stellar mass.

This differs from the calculation in Section~\ref{subsec:rot}, where we include all stellar particles within a fixed physical aperture. There, our aim is to follow the build-up of angular momentum throughout the formation and assembly of the massive GC candidate, when the system is still evolving and its bound stellar component is not necessarily well defined.

\section{Extended data}
\label{sec:exdata}

\begin{figure}[h]
\centering
\includegraphics[width=1\hsize]{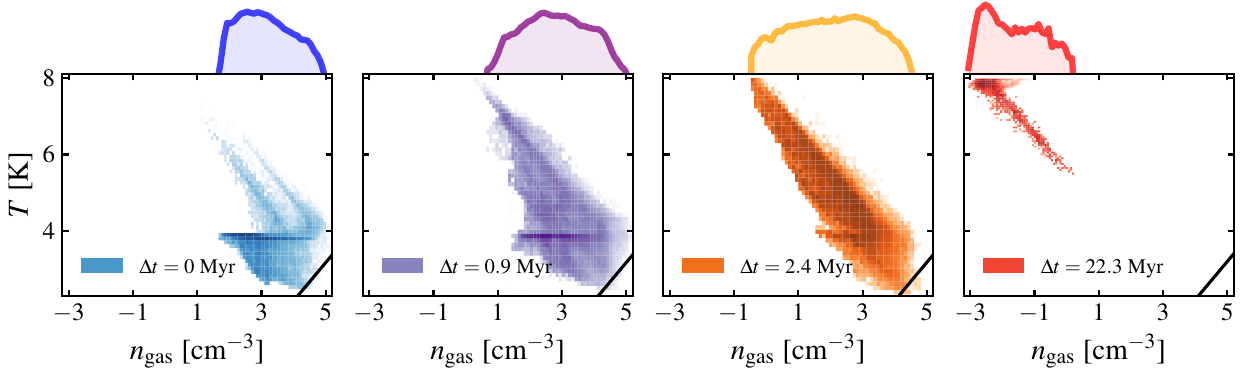}
    \caption{Phase diagram of gas during the formation of the GC candidate. The phase diagrams are constructed from a cubic region of side $15\pc$ centred on the GC candidate. The panels span the evolution of a $22.3\Myr$ baseline, the same time span followed in Fig.~\ref{fig:fig1}. The black line at the bottom right shows the line below which Jeans length is not resolved anymore (the Truelove criterion \cite{Truelove1997}). The line is constructed imposing that the Jeans length is resolved by at least 4 gas cells.}\label{fig:phase}
\end{figure}

\begin{figure}
    \centering
    \includegraphics[width=1\hsize]{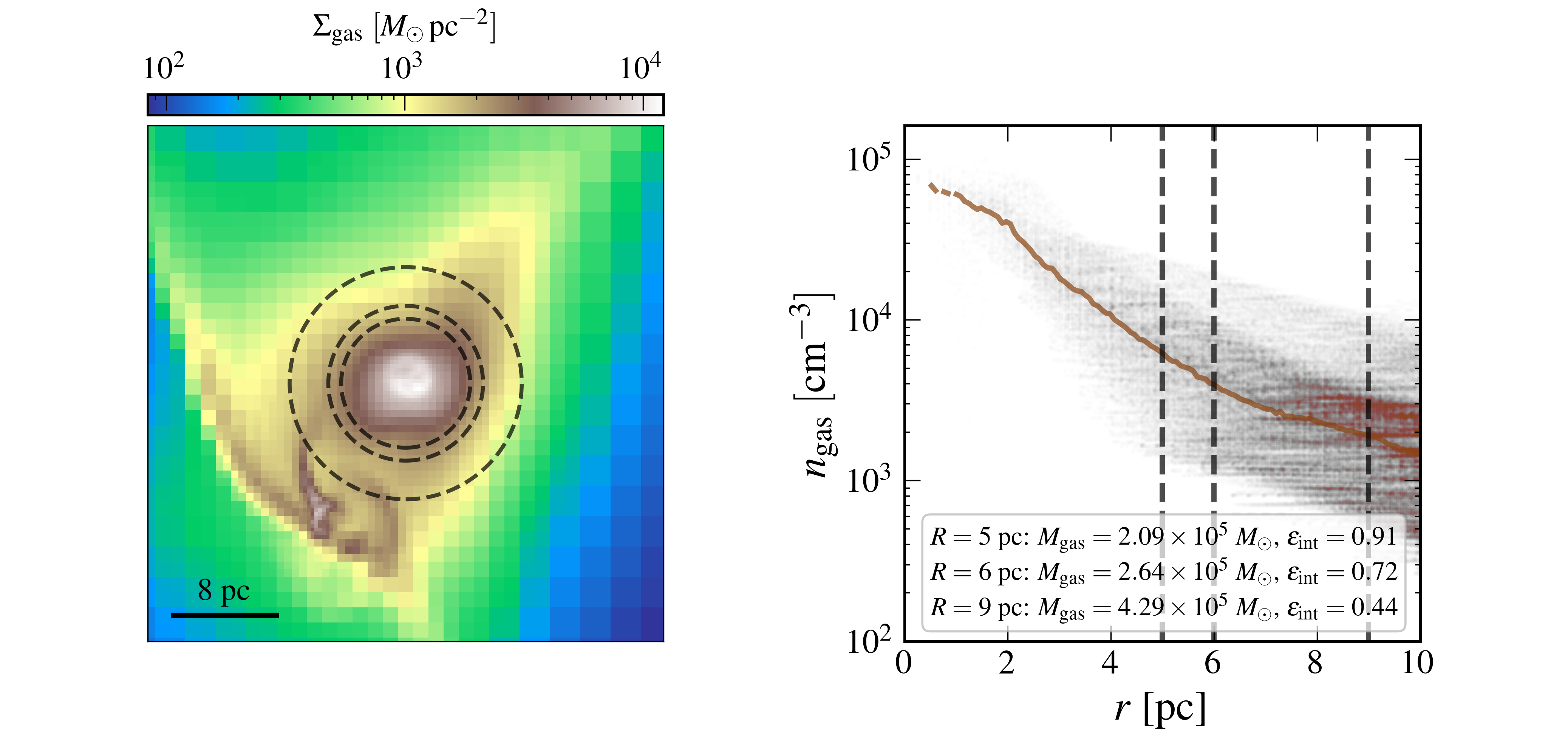}  
    \caption{Integrated star-formation efficiency of the GC candidate. Left panel: Projected gas-density distribution around the GC candidate at the onset of star formation, $\Delta t=0$. The circles mark radii of 5, 6, and 9 pc from the centre. 5 pc is the distance from the center where the central density has decreased by ten \citep{Garcia2023}. Right panel: Three-dimensional gas-density profile at the same time. Each point shows the density of an individual gas cell, while the solid line indicates the median profile. Vertical dashed lines mark the same radii shown in the left panel. The corresponding integrated star-formation efficiencies are $\epsilon_{\rm int}=0.91$, 0.72, and 0.44, respectively, computed using the stellar mass formed after 3 Myr, $1.9\times10^5\Msun$.}
        \label{fig:sfe}
\end{figure}

\clearpage

\section*{Supplementary Information}

\subsection*{Gas properties and numerical resolution}
\label{subsec:gp}

To further characterise the gas evolution during the formation of the system, we analyse the density distribution and thermodynamical state of the gas in a cubic region of side $15\pc$ centred on the GC candidate at different stages of its evolution (Extended Data Fig.~\ref{fig:phase}). At the onset of star formation, the gas is predominantly concentrated at high densities, with $n \gtrsim 10^3\cm^{-3}$. As the system evolves, the density distribution progressively shifts towards lower values and develops an extended low-density tail. In the final stages, the gas is largely rarefied, with densities reaching values as low as $n \simeq10^{-3}\cm^{-3}$, while the high-density component is almost entirely depleted. The phase diagrams shown in Fig.~\ref{fig:phase} indicate that, at early times, the dense gas is predominantly cold and lies below the temperature threshold required to trigger star formation in the model. As stellar feedback becomes effective, a distinct hot and diffuse component emerges at lower densities. This transition reflects the impact of early feedback, primarily driven by stellar winds, which efficiently heat the gas and reduce its density, contributing to the rapid quenching of star formation. 

To quantify how efficiently the initial gas reservoir is converted into stars, we also estimated the integrated star-formation efficiency, $\epsilon_{\rm int}$, defined as the fraction of the initial gas mass of the cloud that is converted into stars. This quantity is related to, but distinct from, the star formation efficiency per free-fall time $\epsilon_{\rm SFE}=1$ adopted in our star-formation prescription (equation~\ref{for:sfe}); although the two cannot be directly equated, a higher $\epsilon_{\rm SFE}$ is generally expected to result in a higher integrated efficiency, but $\epsilon_{\rm int}$ provides a more direct comparison with efficiencies inferred from observations and simulations \cite{Grudic2019,Polak2024,Turner2015,RicoVillas2020}. We estimate the initial cloud mass at the onset of star formation ($\Delta t=0$, see Extended Data Figure \ref{fig:sfe}). Following \cite{Garcia2023}, we define the approximate edge of the natal cloud within which the GC candidate forms as the radius where the initial gas density decreases by one order of magnitude from its central value \citep{Garcia2023}, giving $R\simeq5$ pc and $\epsilon_{\rm int}\simeq0.91$. Adopting 6 and 9 pc, which better encompass the GC candidate size and account for its partial assembly from gas originating in adjacent star-forming regions, yields $\epsilon_{\rm int}\simeq0.72$ and $0.44$, respectively. These values should therefore be regarded as indicative, but remain high and consistent with those found in dense massive-cluster-forming environments \citep{Grudic2019,Polak2024,Turner2015}.

We also assess the numerical resolution of the gas collapse by evaluating the Jeans condition, requiring the Jeans length to be resolved by at least four resolution elements \cite{Truelove1997}. Only a small fraction of the gas mass ($\simeq 4 \times 10^3\Msun$, corresponding to $\sim 1\%$ of the total stellar mass formed in the clump) falls below this resolution limit. This indicates that the collapse of the system is well resolved and not significantly affected by artificial fragmentation \citep{Truelove1997}.

\backmatter

\bmhead{Acknowledgements}
RP acknowledges support for this study from the INAF Mini Grant 2025 (Ob.Fu.1.05.24.07.05, CUP C33C24001390005). RP also acknowledges funding from the INAF Theory 2024 Grant project ``Magnetohydrodynamic Simulations of Galactic Molecular Clouds: Resolving Stellar Birth and Proto-planetary Discs with an Enhanced Chemical Network''. This research was supported in part by Lilly Endowment, Inc., through its support for the Indiana
University Pervasive Technology Institute.

\bmhead{Correspondence}
Correspondence and requests for materials should be addressed to R.P.

\bmhead{Author contributions}
R.P. carried out the analysis of the simulations and wrote the manuscript. F.C. performed the simulations and supervised the project. E.V., A.D.C., J.R., and A.L. contributed to the interpretation and discussion of the results.

\bmhead{Competing Interests}
The authors declare no competing interests.

\bibliography{bibliography}

\end{document}